\documentclass[preprint,final,5p,times,twocolumn]{elsarticle}
\usepackage{rotating,color,subfigure,amssymb}
\usepackage{alphalph}\renewcommand\theaffn{\alphalph{\arabic{affn}}}
\usepackage{amsmath}
\usepackage[T1]{fontenc}
\usepackage{amssymb}
\journal{Physics Letters B}
\begin{document}
                                       
\begin{frontmatter}

\title{Deuteron photodisintegration by polarized photons in the region of the $d^*(2380)$}
\date{\today}

\author[UoY]{M.~Bashkanov}\ead{mikhail.bashkanov@york.ac.uk}
\author[UoR]{S.~Kay}
\author[UoY]{D.P.~Watts}
\author[UoG]{C. Mullen}
\author[UBasel]{S.~Abt}
\author[KPM]{P.~Achenbach}
\author[KPM]{P.~Adlarson}
\author[UoB]{F.~Afzal}
\author[UoR]{Z.~Ahmed}
\author[KSU]{C.S.~Akondi}
\author[UoG]{J.R.M.~Annand}
\author[KPM]{H.J.~Arends}
%
\author[UoB]{R.~Beck}
\author[KPM]{M.~Biroth}
\author[Dubna]{N.~Borisov}
\author[INFN]{A.~Braghieri}
\author[GWU]{W.J.~Briscoe}
%
\author[KPM]{F.~Cividini}
\author[SMU]{C.~Collicott}
\author[Pavia,INFN]{S.~Costanza}
%
\author[KPM]{A.~Denig}
\author[GWU]{E.J.~Downie}
\author[Giessen,KPM]{P.~Drexler}
%
%
\author[UBasel]{S.~Garni}
\author[UoG]{D.I.~Glazier}
\author[Dubna]{I.~Gorodnov}
\author[KPM]{W.~Gradl}
\author[UBasel]{M.~G{\"u}nther}
\author[INR]{D.~Gurevich}
%
\author[KPM]{L. Heijkenskj{\"o}ld}
\author[MAU]{D.~Hornidge}
\author[UoR]{G.M.~Huber}
%
\author[UBasel]{A.~K{\"a}ser}
\author[KPM,Dubna]{V.L.~Kashevarov}

\author[RBI]{M.~Korolija}
\author[UBasel]{B.~Krusche}
%
\author[Dubna]{A.~Lazarev}
\author[UoG]{K.~Livingston}
\author[UBasel]{S.~Lutterer}
%
\author[UoG]{I.J.D.~MacGregor}
\author[UoG]{R. Macrae}
\author[KSU]{D.M.~Manley}
\author[KPM,MAU]{P.P.~Martel}
\author[UoM]{R.~Miskimen}
\author[KPM]{E.~Mornacchi}
%
\author[Dubna]{A.~Neganov}
\author[KPM]{A.~Neiser}
%
\author[KPM]{M.~Ostrick}
\author[KPM]{P.B.~Otte}
%
\author[UoR]{D.~Paudyal}
\author[INFN]{P.~Pedroni}
\author[UoG]{A.~Powell}
\author[UoC]{S.N.~Prakhov}

%
\author[RIP]{G.~Ron}
\author[Basel,RU]{T.~Rostomyan}
%
\author[SMU]{A.~Sarty}
%
\author[KPM]{C.~Sfienti}
\author[KPM]{V.~Sokhoyan}
\author[UoB]{K.~Spieker}
\author[KPM]{O.~Steffen}
\author[GWU]{I.I.~Strakovsky}
\author[UBasel]{T.~Strub}
\author[RBI]{I.~Supek}
%
\author[UoB]{A.~Thiel}
\author[KPM]{M.~Thiel}
\author[KPM]{A.~Thomas}
%
\author[Dubna]{Yu.A.~Usov}
%
\author[KPM]{S.~Wagner}
\author[UBasel]{N.K.~Walford}
\author[UoY]{D.~Werthm\"uller}
\author[KPM]{J.~Wettig}
\author[KPM]{M.~Wolfes}
%
\author[JLab]{L.A.~Zana}

  
\address[UoY]{Department of Physics, University of York, Heslington, York, Y010 5DD, UK}
\address[UoR]{University of Regina, Regina, SK S4S-0A2 Canada}

\address[UoG]{SUPA School of Physics and Astronomy, University of Glasgow, Glasgow, G12 8QQ, UK}
\address[UBasel]{Department of Physics, University of Basel, Ch-4056 Basel, Switzerland}
\address[KPM]{Institut f\"ur Kernphysik, University of Mainz, D-55099 Mainz, Germany}
\address[UoB]{Helmholtz-Institut f\"ur Strahlen- und Kernphysik, University Bonn, D-53115 Bonn, Germany}
\address[KSU]{Kent State University, Kent, Ohio 44242, USA}

\address[Dubna]{Joint Institute for Nuclear Research, 141980 Dubna, Russia}

\address[INFN]{INFN Sezione di Pavia, I-27100 Pavia, Pavia, Italy}

\address[GWU]{Center for Nuclear Studies, The George Washington University, Washington, DC 20052, USA}
\address[SMU]{Department of Astronomy and Physics, Saint Mary's University, E4L1E6 Halifax, Canada}
\address[Pavia]{Dipartimento di Fisica, Universit\`a di Pavia, I-27100 Pavia, Italy}
\address[Giessen]{II. Physikalisches Institut, University of Giessen, D-35392 Giessen, Germany}
\address[INR]{Institute for Nuclear Research, RU-125047 Moscow, Russia}

\address[MAU]{Mount Allison University, Sackville, New Brunswick E4L1E6, Canada}
\address[RBI]{Rudjer Boskovic Institute, HR-10000 Zagreb, Croatia}

\address[UoM]{University of Massachusetts, Amherst, Massachusetts 01003, USA}
\address[UoC]{University of California Los Angeles, Los Angeles, California 90095-1547, USA}
\address[RIP]{Racah Institute of Physics, Hebrew University of Jerusalem, Jerusalem 91904, Israel}
\address[RU]{current address: Department of Physics and Astronomy, Rutgers University, Piscataway, New Jersey, 08854-8019}

\address[JLab]{Jefferson Lab, 12000 Jefferson Ave., Newport News, VA 23606, USA}

\cortext[coau]{Corresponding author }

\begin{abstract}
We report the first large-acceptance measurement of the beam-spin asymmetry for deuteron photodisintegration ($\vec{\gamma} d\to pn$) in the photon energy range $420<E_{\gamma}<620$~MeV. The measurement provides important new constraints on the mechanisms of photodisintegration above the $\Delta$ resonance and on the  photocoupling of the recently discovered $d^*(2380)$ hexaquark.

\end{abstract}

\end{frontmatter}


\section{\label{sec:Int}Introduction}

The deuteron is the simplest nucleus in nature, a bound two-body system comprising a proton-neutron pair. The most elementary nuclear reaction process is photodisintegration, where the deuteron is disintegrated into its component proton and neutron through its interaction with a photon.  At low photon energies (a few to tens of MeV) such reactions play an important role in nucleosynthesis and stellar burning. At higher photon energies ($E_\gamma > 100$~MeV), the process becomes sensitive to the nucleonic and ultimately ($E_\gamma > 1000$~MeV) quark substructures of the deuteron~\cite{Gross}.   

More than 50 years ago, it became clear that deuteron photodisintegration can be useful to understand the nucleon-nucleon and $NN^*$ interactions (where $N^*$ represents an excited state of the nucleon) as well as to search for exotic six-quark particles, the so-called hexaquarks~\cite{Chadwick}. Despite its importance, the world data set for deuteron photodisintegation contains significant gaps in terms of photon energy coverage, angular coverage, and particularly in measurements of polarization observables. The situation has become more critical following the recent exciting indications for the discovery of the first exotic hexaquark - the $d^*(2380)$ with quantum numbers $I(J^P)=0(3^+)$~\cite{mb,MB,MBC,TS1,TS2,MBA,MBE1,MBE2}. With such quantum numbers, the $d^*(2380)$ can be produced on deuterium with photon beams of fairly moderate energy ($E_{\gamma}\approx 570$~MeV)~\cite{BBC}.

The optimal reactions to search for evidence of the $d^*(2380)$ in photoreactions are $\gamma d\to d^*\to d\pi^0\pi^0$ and $\gamma d\to d^* \to pn$, which include the dominant decay branches established in nucleonic beam experiments~\cite{BCS,BCSW}. The isospin selectivity of the $d\pi^0\pi^0$ final state suppresses backgrounds from conventional nucleon resonances leaving the $d^*(2380)$ as the strongest contribution~\cite{BCSW}. Although weaker suppression of background mechanisms is expected for the $pn$ final state, the simpler two-body final state is far more amenable to detailed partial wave analysis.

The $\gamma d\to d\pi^0\pi^0$ reaction was recently measured by the A2 collaboration at MAMI~\cite{Bas} and by the Tohoku Collaboration~\cite{Toh}. Both measurements gave similar results $\sigma (\gamma d\to d^*\to d\pi^0\pi^0)\approx 20 \text{ -- } 30$~nb (on top of $\sigma (\gamma d\to d\pi^0\pi^0) \sim 15$~nb of conventional processes). Due to the small cross section and the closeness of the $d^*(2380)$ to the reaction threshold, a complete determination of the final state, necessary to suppress background in the region of the $d^*(2380)$ and reduce systematics, will require further dedicated measurements~\cite{BDProp}.

The photodisintegration reaction has a higher estimated background level~\cite{BCSW}:
\begin{equation}
  \frac{\sigma(\gamma d\to d^*\to pn)}{\sigma(\gamma d\to pn)}\approx \frac{30~nb}{6~\mu b}\approx 10^{-2}.
\end{equation}
  However, by exploiting polarization observables together with partial-wave analysis, one can reach higher sensitivities to the $d^*(2380)$. The possibility exists not only to extract the $d^*(2380)$ photo-coupling per se, but also to get hints about the $d^*(2380)$ electromagnetic properties, such as its electric quadrupole moment, its magnetic octupole moment, etc.

Indeed, the first irregularities in deuteron photodisintegration were observed in the 1970s in proton polarization measurements $d(\gamma,\vec{p})n$~\cite{TOK1,TOK2}. It was suggested that the high proton polarization ($P_p\sim 100\%$) observed at $E_\gamma \approx 570$~MeV might originate from a yet unknown six-quark particle with mass $\sim 2380$~MeV and quantum numbers $I(J^P)=0(3^+)$, consistent with the $d^*(2380)$ later found in nucleon-nucleon scattering experiments. Due to its high spin, $J^P_{d^*}=3^+$, the $d^*(2380)$ requires the contribution of higher multipoles ($E2, M3$ or $E4$) to be photoproduced from a deuteron target ($I(J^P)=0(1^+)$). Its decay to a proton-neutron pair also requires high partial waves: according to Refs.~\cite{MBE1,MBE2} it proceeds in 90\% of cases via the $^3D_3$ partial wave (angular momentum $L=2$, nucleon spins and $L$ all aligned) or in 10\% of cases via the $^3G_3$ partial wave (angular momentum $L=4$, nucleon spins aligned, spin and $L$ anti-aligned). Such high angular-momentum components are expected to reveal themselves in various polarization observables. If true, the $d^*(2380)$ would be the only known exotic multiquark system that can be produced copiously in the clean and controlled environment of photo-induced reactions, allowing study of the structure of multiquark systems with unprecedented accuracy. To verify this assumption we aim to determine a much more complete measurement of polarization observables for the $\gamma d\to pn$ reaction~\cite{DWProp} and the first step in this direction is a measurement of the $d(\vec{\gamma},pn)$ reaction to determine the beam-spin asymmetry, $\Sigma$, in the region of the $d^*(2380)$. 

\section{\label{sec:app}Experimental details}

This experiment took place at the Mainz Microtron (MAMI) electron accelerator facility~\cite{MAMI} in a total beamtime of 300 hours during August 2016. Bremsstrahlung photons, either circularly or linearly polarized, were energy tagged ($\Delta E\sim 2$~MeV) by the Glasgow-Mainz Tagger~\cite{Tagg} and impinged on a 10 cm long liquid deuterium target. Reaction products were detected by the Crystal Ball (CB)~\cite{CB}, a highly segmented NaI(Tl) photon calorimeter covering nearly 96\% of $4\pi$ ($21^{\circ}< \Theta <159^{\circ}$). A 24 element, 30 cm long plastic scintillator barrel (PID) surrounded the target to assist in charged particle identification~\cite{PID}. For this experiment, additional analyzing material for a nucleon polarimeter was placed inside the CB, comprising a 2.6 cm thick graphite cylinder covering $\Theta > 12^{\circ}$ placed in the space between the PID and the Multi Wire Proportional Chamber (MWPC)~\cite{MWPC}. A further component of the polarimeter, a 2.6 cm thick disc-shaped upstream cap covered $2^{\circ}< \Theta < 12^{\circ}$~\cite{PID}. The polarimeter was not used for this first analysis but will allow nucleon polarization observables to be extracted from the same data set in subsequent analysis (ongoing).

The $d(\vec{\gamma},pn)$ events of interest contained a proton track and an uncharged (neutron) hit. The proton was identified using the correlation between the energy deposits in the PID and CB using $\Delta E$ - $E$ analysis~\cite{PID}. The proton identification also required an associated charged track in the MWPC. The neutron candidates comprised uncharged hits in the CB, which did not have any associated MWPC or PID signal. The neutron angles were determined using the CB hit with production vertex coordinates defined by the intercept of the photon beam trajectory and the associated charged (proton) track. Once candidate proton and neutron tracks were identified, a kinematic fit was employed to increase the purity of the event sample and to improve the accuracy in the determination of the reaction kinematics. To fully constrain the kinematics of the $d(\gamma,pn)$ reaction at a given incident photon energy, two kinematic quantities are required. The angles of both proton and neutron as well as the proton energy are measured, enabling an overconstrained kinematic fit analysis. A 10\% probability cut was employed to select the events of interest. To remove events originating from the target cell windows, a $z$-vertex cut was also imposed. The final data sample contains $\sim 5\times 10^6$ events in the range of interest ($E_{\gamma}\sim 420-620$~MeV). The neutron detection efficiency of the CB ($\sim$~30\%) was also determined in a dedicated measurement, Ref.~\cite{NEff}. Applying these detection efficiencies in the current analysis enabled differential cross sections to be extracted, which were consistent with previously measured DAPHNE data~\cite{DAPHNE}.

\begin{figure}[!h]
\begin{center}
\includegraphics[width=0.4\textwidth,angle=0]{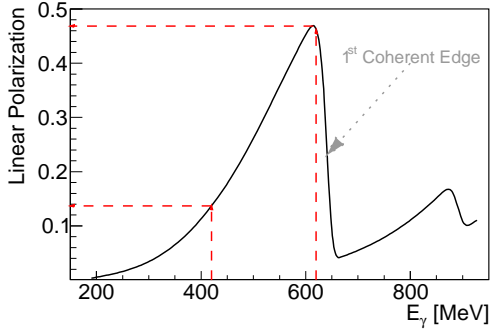}
\end{center}
\caption{The linear polarization of the bremsstrahlung photon beam as a function of photon energy. The coherent edge marks the point where the second derivative of the polarization as a function of the energy changes in sign. The vertical (horizontal) dashed lines shows accessible range of energies (polarizations) for this experiment.}
\label{GPolFig}
\end{figure}

A linearly polarized photon beam was employed for 2/3 of the experimental run period, with the remainder obtained using an unpolarized\footnote{Circularly polarized with equal number of events from both helicities.} photon beam. The linearly polarized beam was produced by utilizing a crystalline diamond radiator to produce the bremsstrahlung photons from the electron beam~\cite{GPol,GPolK}. Systematics in the measurement were reduced by periodically switching the plane of the linear polarization by $90^{\circ}$. Also, the unpolarized data were obtained at regular intervals throughout the run period. The coherent edge for the linearly polarized photons was set at 630 MeV, which provided linearly polarized photons with appreciable polarization  ($14\%<P_{\gamma}<47\%$) over the energy range $E_{\gamma}\sim 420-620$~MeV, see Fig.~\ref{GPolFig}.

\section{\label{sec:sigma} Determination of the beam-spin asymmetry}

The differential cross section for deuteron photodisintegration is related to the beam-spin asymmetry, $\Sigma$, by
\begin{equation}
  \frac{d\sigma}{d\Omega}=\bigg( \frac{d\sigma}{d\Omega} \bigg)_{unpol} (1+P_\gamma\Sigma\cdot \cos(2\phi)) ,
\end{equation}
  where $P_\gamma$ is the degree of linear polarization of the photon beam and $\phi$ is the azimuthal angle between the photon polarization vector and the reaction plane~\cite{NZ}. For this experiment, the polarization plane of the incident photon beam was periodically flipped. The two orientations corresponded to the polarization plane being parallel ($\parallel$) or perpendicular ($\perp$) to the laboratory floor. Such an arrangement allows the beam-spin asymmetry to be determined from a double ratio of the yields in the two polarization directions, which is less sensitive to any systematical effects arising from detector acceptance:

\begin{equation}
P_{\gamma}\Sigma\cdot \cos(2\phi)=\frac{N(\Theta,\phi)_{\parallel}-N(\Theta,\phi)_{\perp}}{N(\Theta,\phi)_{\parallel}+N(\Theta,\phi)_{\perp}}
\end{equation}
    
As a cross check, the asymmetry, $\Sigma$, was also calculated separately from the $\parallel$ and $\perp$ data, using the unpolarized data to remove the effects of detector acceptance. This method gave broadly consistent results with those extracted from the double ratio, with the small (order 3\%) discrepancies included in the quoted systematic uncertainty.

Further systematics were assessed and quantified in the data analysis. This included comparison of the results obtained when not employing a kinematic fit analysis, variation of the probability cut (and also the measured quantities employed) in the kinematic fit,  variation of the range of the proton missing-mass cut\footnote{For the case where the analysis was performed without using kinematic fit techniques, e.g. see Ref~\cite{PID}.}, and variation of the $\Delta E$ - $E$ and vertex cuts.  Little sensitivity to the extracted values of $\Sigma$ was observed (typically changes of below 5\%).  The estimated systematic uncertainties were assessed on a bin-by-bin basis and are presented with our results. An additional 3\% systematic (not included in the bin-by-bin systematics in the figure) arises from the method used to determine the degree of photon beam polarization.

\begin{figure}[!h]
\begin{center}
\includegraphics[width=0.5\textwidth,angle=0]{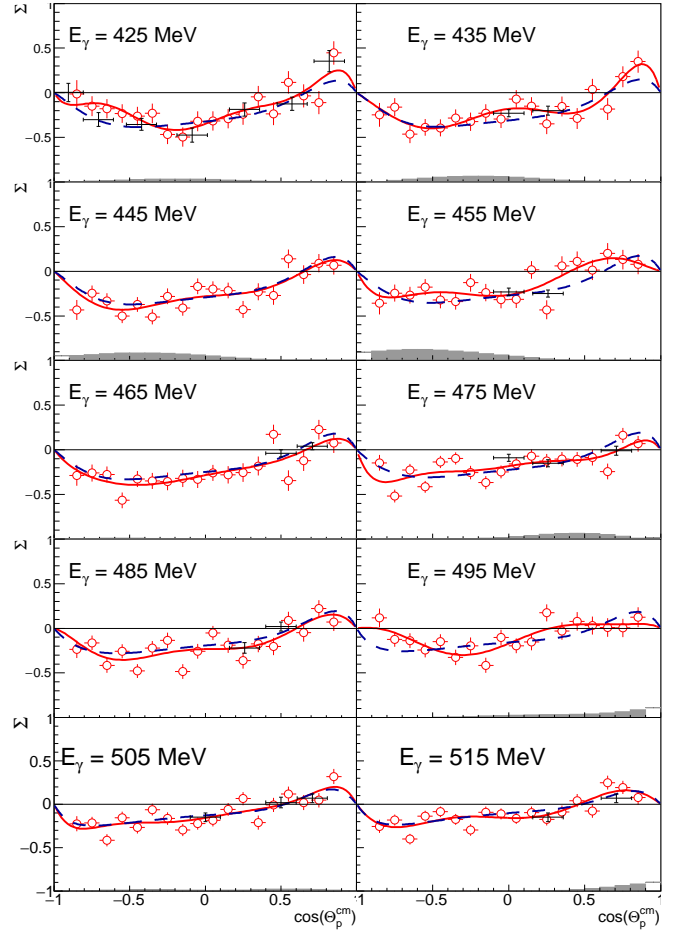}
\end{center}
\caption{(Color online) Beam-spin asymmetry ($\Sigma$) results from this experiment (red open circles) in comparison with previous results (black crosses)~\cite{sig1,sig2,sig3}. The corresponding systematic uncertainties are depicted as shaded bars on the bottom. Energy independent (energy dependent) $a_lP_l^2$ fits are shown as solid red (dashed blue) lines (see text).}
\label{Sig1}
\end{figure}

\section{\label{sec:sigma} Results}

\begin{figure}[!h]
\begin{center}
\includegraphics[width=0.5\textwidth,angle=0]{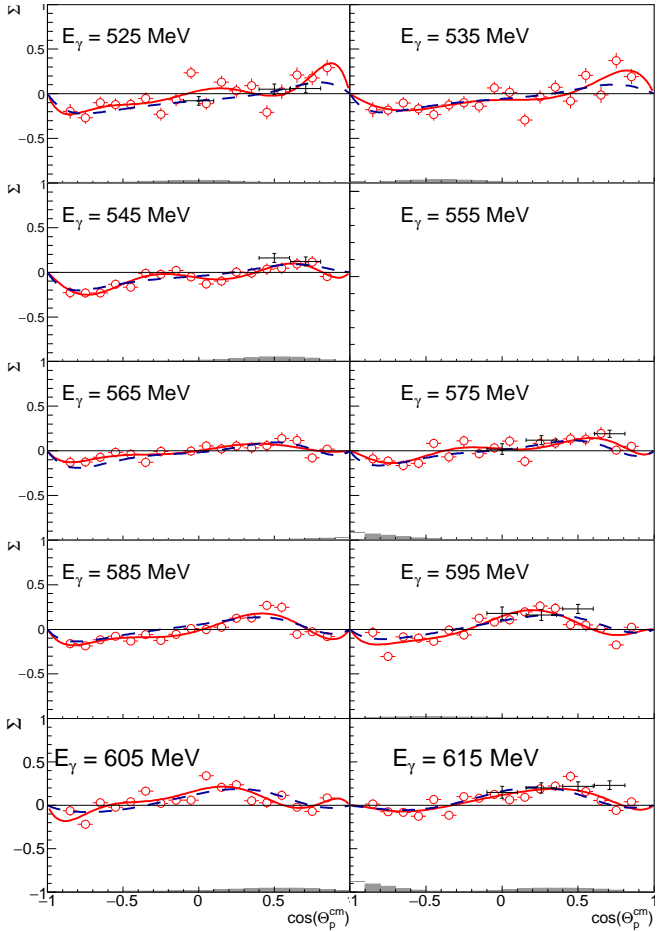}
\end{center}
\caption{Same as Fig.~\ref{Sig1}. The 540-550 MeV bin is empty due to a dead tagger channel.}
\label{Sig2}
\end{figure}

Our beam asymmetry results and estimated systematic uncertainties are presented in Figs.~\ref{Sig1} and~\ref{Sig2}. The new data cover the complete range of proton center-of-mass (CM) polar angles for photon energy bins from 420 up to 620 MeV, providing the first comprehensive measurement of this observable for this photon energy range. Also shown in Figs.~\ref{Sig1} and~\ref{Sig2}. are the previous world experimental data~\cite{sig1,sig2,sig3}. For the lowest $E_\gamma= 425$~MeV bin where there are existing data~\cite{sig1}, consistent results were obtained over the full range of proton angles. Consistency with previous data sets is also observed with the sparse data points at higher energies, which were obtained in restricted kinematics. 

This work provides the first kinematically complete measurement of a polarization observable in deuterium photodisintegration above 420 MeV. The new data will provide stringent tests of any theory of photodisintegration in this region and, importantly, provide the first extensive measurement of a polarization observable in the region of the $d^*(2380)$.

\begin{figure}[!h]
\begin{center}
\includegraphics[width=0.5\textwidth,angle=0]{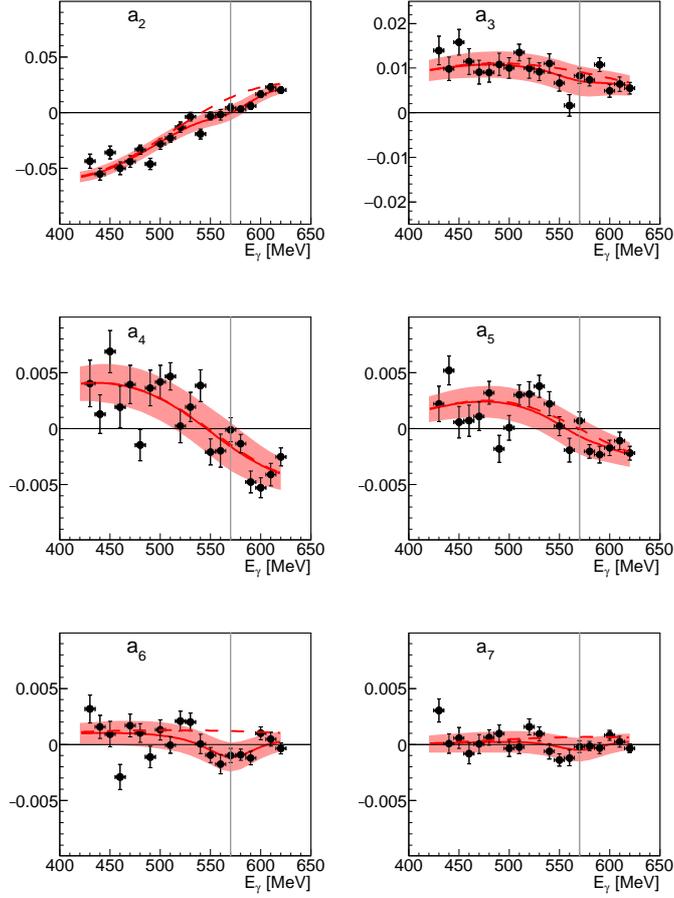}
\end{center}
\caption{The energy dependence of the expansion coefficients $a_l$. Black dots correspond to the single-energy solution, red line correspond to the energy-dependent solution, and red bands represent the $1\sigma$ error band for the energy-dependent solution. The thin vertical grey lines point to the $d^*(2380)$ energy. Dashed red lines show results without a $d^*(2380)$ Breit-Wigner contribution.}
\label{fit1}
\end{figure}

To quantify the dependence of the asymmetry on photon energy and polar angle we performed an expansion of our results into associated Legendre functions. From a theoretical point of view, it has been suggested to decompose not $\Sigma(\Theta,E_\gamma)$, but rather $\sigma_1=\Sigma(\Theta,E_\gamma)\cdot\sigma(\Theta,E_\gamma)$~\cite{Sdec} or even $\sigma_1/\sigma_{\textrm{tot}}$, where $\sigma_{\textrm{tot}}$ is the total cross section, to get rid of the $s^{-10}$ energy dependence of the deuteron photodisintegration cross section. We therefore adopt the  $\sigma_1/\sigma_{\textrm{tot}}$ ansatz, employing the differential cross section and total cross section measurements from Ref.~\cite{DAPHNE}\footnote{Note that the cross sections extracted from the current data were consistent with these previous measurements.} \footnote{Since both $d\sigma/d\Omega$ and $\sigma_{\textrm{tot}}$ cross sections are taken from the same Ref.~\cite{DAPHNE} data, the systematic uncertainties related to the $\sigma_{\textrm{tot}}$ extraction are expected to be suppressed.}. The resulting $\sigma_1/\sigma_{\textrm{tot}}$ data were fitted using the expansion
\begin{equation}
  \frac{\sigma_1}{\sigma_{\textrm{tot}}}=\sum\limits_{l=2}^7 {a_lP_l^2}.
  \end{equation}
This procedure was carried out using two methods: (i) a single-energy procedure in which the fit was performed using data from each photon energy bin in isolation and (ii) an energy-dependent procedure where the expansion coefficients, $a_l$, were assumed to vary smoothly\footnote{To ensure smoothness we have used a Discrete Gaussian Sampling method to model a smooth $a_l$ energy dependence. This comprised 3 Gaussians with centroids at 420, 520 and 620 MeV, widths of 100 MeV with arbitrary strength.} from photon energy bin to photon energy bin. To evaluate the possible influence of the $d^*(2380)$ to the photon energy dependence of the $a_l$ coefficients we also inserted a Breit-Wigner function having a mass $M = 2380$~MeV, a width $\Gamma = 80$~MeV, and arbitrary strength for each $a_l$. The extracted coefficients using the various ansatz, with and without the modelling of the $d^*(2380)$ contribution, are presented as a function of photon energy in Fig.~\ref{fit1}. 

In Ref.~\cite{Ikeda}, it is claimed that the strongest $d^*(2380)$ effect should be seen in the $a_6$ coefficient. Some weak structure exhibiting a mass and width compatible with the established properties of the $d^*(2380)$ is evident in the $a_6$ coefficient. A $d^*(2380)$ contribution from $E2$ in a product with higher multipoles ($E4$ or higher) is allowed; however, it is expected to be suppressed~\cite{Arn}. The observed rather smooth behavior of $a_4$, where an $E2$ contribution from the $d^*(2380)$ may be expected to manifest, is consistent with a very weak $E2$ excitation of the deuteron into $d^*(2380)$, in favor  of $M3$ excitation, which should manifest itself in $a_6$, but not in $a_4$. This result may not be as unexpected as it looks at first sight: due to the predicted compactness of the $d^*(2380)$~\cite{DongFF} its quadrupole deformation and hence the deuteron to $d^*$ electric quadrupole transition is expected to be small\footnote{The leading contribution in an $E2$ transition is expected to arise either from $D-$wave (deuteron) to $S-$wave $\Delta\Delta$ ($d^*(2380)$) or $S-$wave (deuteron) to $D-$wave $\Delta\Delta$ ($d^*(2380)$). The transition from the $pn$ part of the deuteron wave function to the six-quark part of the $d^*$ wave function is prohibited because the photon does not carry color charge and cannot change two color bags into one. The transition from a six-quark part of the deuteron to a six-quark part of the $d^*$ wave function via an $E2$ transition should be highly suppressed.}. On the other hand, the magnetic moment of the $d^*(2380)$ is expected to be large, $\mu_{d^{*}}\approx 7.6~\mu_B$~\cite{DongFF}, which should lead to an enhanced magnetic octupole transition. The $a_6$ coefficient appears to be the last nontrivial coefficient for the sampled photon energy range. The $a_7$ coefficient is consistent with zero within our statistical and systematic uncertainties.

One can also use the formalism from Ref.~\cite{Well92} to evaluate the anticipated correlations and signal sizes expected in the various legendre coefficients. Assuming the $d^*(2380)$ is excited via the $M3$ transition only and neglecting the interference effects between the $d^*(2380)$ and $NN^*$ -systems (no $d^*(2380)$ - $NN^*$ interference has been previously observed~\cite{Bas, MB, MBE1,MBE2,TS2}), we can write simple expressions for the $a_l$ coefficients (the $d^*(2380)$ contribution to the odd $a_l$ coefficients appear only as an interference with other partial waves are therefore expected to be small).

\begin{eqnarray}  
  a_2=&0.375|M3(^3D_3)|^2+0.108|M3(^3D_3)||M3(^3G_3)| \nonumber\\
  &\cdot\cos(\delta_{^3D_3}-\delta_{^3G_3})+0.391|M3(^3G_3)|^2 \nonumber\\
  \nonumber\\
  a_4=&0.014|M3(^3D_3)|^2+0.021|M3(^3D_3)||M3(^3G_3)| \nonumber\\
  &\cdot\cos(\delta_{^3D_3}-\delta_{^3G_3})+0.017|M3(^3G_3)|^2 \nonumber\\
  \nonumber\\
  a_6=&~~~~~~~~~~~~~~~~~~~~~~0.172|M3(^3D_3)||M3(^3G_3)| \nonumber\\
  &\cdot\cos(\delta_{^3D_3}-\delta_{^3G_3})+0.025|M3(^3G_3)|^2  
\end{eqnarray}

where the first number corresponds to the multipole transition in the initial state and the numbers in brackets correspond to the $pn$ partial wave in conventional nomenclature $(^{2S+1}L_J)$. The relative phases between the $^3D_3$ and $^3G_3$ partial waves are determined to be $\sim 4^{\circ}$  from the AD14 SAID partial wave analysis~\cite{MBE2,AD14}. Using $\frac{|d^*(^3D_3)|^2}{|d^*(^3G_3)|^2}=9$ from the same analysis we can simplify the expression:

\begin{eqnarray}
  a_2=0.454|M3(^3D_3)|^2, \nonumber\\
  a_4=0.023|M3(^3D_3)|^2, \nonumber\\
  a_6=0.060|M3(^3D_3)|^2. 
\end{eqnarray}

In table~\ref{tab:1} we show results where the $a_6$ strength is fixed from the data and use the relations (6) to estimate the expected strength for other coefficients based on our previous assumptions. These estimates are  consistent with  the experimental data for both the $a_2$ and $a_4$ coefficients, giving further indication that the the features in the data are broadly consistent with that expected from a $d^*(2380)$ contribution reached predominantly via $M3$ excitation. If we explicitly impose the relations (6) constrains in the fit ("constrained fit",Tab.~\ref{tab:1}), we still observe very similar results for the coefficients $a_2-a_6$. Similarly one can evaluate the $d^*$ contribution to the total $\gamma d\to pn$ cross-section using the measured coefficients. The obtained value of $4.6\pm 1.0\%$ is of similar magnitude to the preliminary value evaluated in Ref.~\cite{Bas}.

\begin{table}
\caption{$d^*(2380)$ contribution to $a_l$}
\label{tab:1}       

\begin{tabular}{llll}
\hline\noalign{\smallskip}
$a_l$ & fit$\times 1000$ & calculated$\times 1000$ & constrained fit$\times 1000$\\
\noalign{\smallskip}\hline\noalign{\smallskip}

$a_2$ & $-13.8\pm 1.6$ &  $-17.6\pm 2.3$ & $-11.7\pm 2.2$\\
$a_4$ & $-~~0.2\pm 0.5$ &  $-~~0.9\pm 0.2$ & $-~~0.6\pm 0.1$ \\
$a_6$ & $-~~2.3\pm 0.3$ &  $-~~2.3\pm 0.3$ & $-~~1.7\pm 0.3$\\
\noalign{\smallskip}\hline\noalign{\smallskip}
$\chi^2/edf$ & ~~~~2.35 &  & ~~~~2.31\\

\noalign{\smallskip}\hline
\end{tabular}
\end{table}

\section{\label{sec:final} Summary}

New precise large acceptance data on beam-spin asymmetry in deuteron photodisintegration have been obtained covering the photon energy region $420<E_\gamma<620$ MeV. The data will be valuable to evaluate the mechanisms of photodisintegration above the $\Delta$ resonance, providing strong constraints on the role of higher resonances than the $\Delta$ accessible in this region ($N^*(1440)$, $N^*(1520)$, $N^*(1535)$), as well as $NN^*$ interactions. These data provide the first comprehensive measurement of a polarization observable in the region of the $d^*(2380)$. The new data were combined with existing cross-section data to perform a simplified multipole analysis. The simplified analysis indicated that the $d^*(2380)$ is likely to be excited predominantly through an $M3$ transition rather than an $E2$ transition, which is consistent with its proposed compact nature. Upcoming polarization data on $P_y, C_x, O_x$ would allow a partial-wave analysis to be performed that would give further powerful constraints on the possible influence of the $d^*(2380)$ hexaquark on deuteron photodisintegration. We hope this work will encourage further developments for including resonances above the $\Delta$ to enable their study in a well understood few-body system.


\section{Acknowledgement}
This work has been supported by the U.K. STFC (ST/L00478X/1, ST/L005824/1, 57071/1, 50727/1 ) grants, the Deutsche Forschungsgemeinschaft (SFB443, SFB/TR16, and SFB1044), DFG-RFBR (Grant No. 09-02-91330), the European Community Research Infrastructure Activity under the FP6 "Structuring the European Research Area" program (Hadron Physics, Contract No. RII3-CT-2004-506078), Schweizerischer Nationalfonds (Contracts No. 200020-156983, No. 132799, No. 121781, No. 117601, No. 113511), the U.S. Department of Energy (Offices of Science and Nuclear Physics, Awards No. DE-SC0014323, DEFG02-99-ER41110, No. DE-FG02-88ER40415, No. DEFG02-01-ER41194) and National Science Foundation (Grants No. PHY-1039130, No. IIA-1358175), INFN (Italy), and NSERC of Canada (Grant No. SAPPJ2015-00023).



\begin{thebibliography}{9}

\bibitem{Gross}  R. Gilman, F. Gross, J. Phys. G {\bf 28}, R37, (2002).

\bibitem{Chadwick} J. Chadwick,  M. Goldhaber, Nature {\bf 134}, 237, (1934).
  
\bibitem{mb} M. Bashkanov {\it et al.}, Phys. Rev. Lett. {\bf 102}, 052301, (2009).
\bibitem{MB} P. Adlarson {\it et al.}, Phys. Rev. Lett. {\bf 106}, 242302, (2011).
\bibitem{MBC} P. Adlarson {\it et al.}, Phys. Lett. B {\bf 721}, 229, (2013).
\bibitem{TS1} P. Adlarson {\it et al.}, Phys. Rev. C {\bf 88}, 055208, (2013).

\bibitem {TS2} P. Adlarson {\it et al.}, Phys. Lett. B {\bf 743}, 325, (2015).

\bibitem {MBA} P. Adlarson {\it et al.}, Eur. Phys. J. A {\bf 52}, 147, (2016).

\bibitem {MBE1} P. Adlarson {\it et al.}, Phys. Rev. Lett. {\bf 112}, 202301, (2014).

\bibitem {MBE2} P. Adlarson {\it et al.}, Phys. Rev. C {\bf 90}, 035204, (2014).


\bibitem {BBC} M. Bashkanov, S. Brodsky and H. Clement, Phys. Lett. B {\bf 727}, 438, (2013).

\bibitem{BCS} M. Bashkanov, H. Clement and T. Skorodko, Eur. Phys. J. A {\bf 51}, 7, 87, (2015).

\bibitem{BCSW}  M. Bashkanov, H. Clement, T. Skorodko and D.P. Watts, Int. J. Mod. Phys. Conf. Ser. {\bf 46}, 1860033, (2018).

\bibitem{Bas} M. Guenther, Master Thesis, University of Basel (2015); PoS (Hadron2017) 051.
  
\bibitem {Toh}  T. Ishikawa {\it et al.},  Phys. Lett. B  {\bf 772}, 398, (2017).

\bibitem {BDProp}  M. Bashkanov, D.P Watts, An active deuteron target programme with the Crystal Ball at MAMI..  

\bibitem {TOK1} H. Ikeda {\it et al.}, Phys. Rev. Lett. {\bf 42}, 1321, (1979).

\bibitem {TOK2} T. Kamae, T. Fujita, Phys. Rev. Lett. {\bf 38}, 471 (1977).

 \bibitem {DWProp} D.P. Watts, D. Glazier, J. Annand and M. Bashkanov Proposal for Experiment: Polarization observables using the Nucleon polarimeter with the Crystal Ball at MAMI. Internal Publication.  
 

\bibitem {MAMI}  K.-H. Kaiser {\it et al.}, Nucl. Instr. Meth. A {\bf 593}, 159 (2008).
\bibitem {Tagg} J.C. McGeorge {\it et al.}, Eur. Phys. J. A {\bf 37}, 129 (2008).
\bibitem {CB} A. Starostin {\it et al.}, Phys. Rev. C {\bf 64}, 055205 (2001).
  
\bibitem {PID} S.J.D. Kay, Ph.D. thesis, University of Edinburgh, (2018).

 \bibitem {MWPC} G. Audit {\it et al.} Nucl. Instr. Meth. A {\bf 301}, 473, (1991).
  
\bibitem {NEff} M. Martemianov {\it et al.} JINST {\bf 10}, no.04, T04001, (2015). 

 \bibitem {DAPHNE} R. Crawford {\it et al.} Nucl. Phys. A {\bf 603}, 303-325, (1996). 

\bibitem {GPol} D. Lohmann {\it et al.}, Nucl. Instrum. Methods Phys. Res., Sect. A {\bf 343}, 494, (1994).

\bibitem {GPolK}  K. Livingston, Nucl. Instr. Meth. A {\bf 603}, 205, (2009). 

\bibitem {NZ} N. Zachariou {\it et al.} Phys. Rev. C {\bf 91}, 055202, (2015). 


\bibitem {sig1} F. V. Adamian {\it et al.} Jour. Phys. G 17(8):1189, (1991).

\bibitem {sig2} V.G. Gorbenko  {\it et al.} Nucl. Phys. A {\bf 381}, 330, (1982).
\bibitem {sig3} S. Wartenberg {\it et al.} Few-Body Systems {\bf 26}, 213 (1999).

\bibitem {Sdec} A. Cambi, B. Mosconi, and P. Ricci Phys. Rev. C {\bf 26}, 2358, (1982).

\bibitem {Ikeda} H. Ikeda {\it et al.}, Nucl. Phys. B {\bf 172}, 509, (1980).

\bibitem {Arn} H. Arenh\"ovel, M. Sanzone, "Photodisintegration of the Deuteron: A Review of Theory and Experiment", ISBN 978-3-7091-6701-4.
  
\bibitem {DongFF} Y. Dong,  P. Shen, Z. Zhang, Phys. Rev. D {\bf 97}, 114002, (2018).
  
\bibitem {Well92} H.R. Weller {\it et al.} Atomic Data and Nuclear Data Tables, {\bf 50}, (1992), 29

\bibitem {AD14} R.L. Workman, W.J. Briscoe, I.I. Strakovsky, Phys. Rev. C {\bf 93}, 045201, (2016). 






\end{thebibliography}
\end{document}